\documentclass[twocolumn,showpacs,preprintnumbers,amsmath,amssymb]{revtex4}
\usepackage{latexsym}
\usepackage{graphicx}
\usepackage{dcolumn}
\usepackage{bm}
\usepackage{xcolor}
\usepackage{float}

\input amssym.def
\input amssym.tex

\newcommand{\mZ}{{\mathbb Z}}
\newcommand{\mN}{{\mathbb N}}

\newcommand{\half}{{{\textstyle\frac{1}{2}}}}

\newcommand{\be}{\begin{equation}}
\newcommand{\ee}{\end{equation} }
\newcommand{\beqa}{\begin{eqnarray} }
\newcommand{\eeqa}{\end{eqnarray} }
\newcommand{\ba}{\begin{array}}
\newcommand{\ea}{\end{array}}

\newcommand\rd{{\rm d}}

\newcommand\cC{{\cal C}}

\newcommand\cS{{\cal S}}
\newcommand\cT{{\cal T}}
\newcommand\cV{{\cal V}}
\newcommand\cZ{{\cal Z}}

\newcommand\BEC{{\rm\scriptscriptstyle{BEC}}}
\newcommand\kB{k_{{\scriptscriptstyle{\rm B}}}}

\newcommand\dis{\displaystyle}

\newcommand\Tp{{\cT}_{{\scriptscriptstyle{P}}}}
\newcommand\Vp{{\cV}_{{\scriptscriptstyle{P}}}}
\newcommand\Trho{{\cT}_{{\rho}}}

\newcommand\vn{\vec{n}}
\newcommand\ve{\varepsilon}
\newcommand\NAv{N_{\scriptstyle\rm{{A}}}}

\begin{document}
\title{Isobar of an ideal Bose gas within the grand canonical ensemble}

\author{Imtak Jeon${}^{\dagger}$, Sang-Woo Kim${}^{\natural}$  and Jeong-Hyuck Park${}^{\dagger}$\footnote{Authors are listed in alphabetical order and any correspondence  should be addressed to {\rm{park@sogang.ac.kr}}}}

\affiliation{{~}\\
\mbox{${}^{\dagger}$Department of Physics,  Sogang University, Seoul 121-742, Korea}\\
\mbox{${}^{\natural}$Department of Physics, Osaka University, Toyonaka, Osaka 560-0043, Japan}}


\begin{abstract}
We investigate    the  isobar of an ideal Bose gas confined in a cubic box within the grand canonical ensemble, for a large yet  finite  number of particles,  $N$. After solving  the equation of the spinodal curve, we derive  precise formulae for the supercooling and the superheating  temperatures which reveal  an $N^{-{1/3}}$ or $N^{-{1/4}}$ power  correction to the known Bose-Einstein condensation temperature in the thermodynamic limit.   Numerical computations confirm the accuracy of our analytical approximation, and further show that the isobar  zigzags on the temperature-volume plane if $N\geq 14393$.  In particular,  for the Avogadro's number of particles,   the volume   expands  discretely about $10^{5}$ times.   Our results  quantitatively  agree with a previous study on the  canonical ensemble within $0.1\%$ error.   
\end{abstract}

\maketitle

\section{Introduction}
A classic paper by Anderson in 1972 goes with the title, \textit{More Is Different} \cite{Anderson72},  which  characterizes the notion of 
`{emergence}':    the way complex systems and patterns arise out of a multiplicity of relatively simple interactions. One relevant  question is then, \textit{How many is different?} To answer the question, we may consult quantum statistical physics where   the key quantity  is  the partition function. Once we know the exact expression   of the partition function,  we can compute various  physical quantities.  For example, when the partition function in the grand canonical ensemble, $\cZ(T,V,z)$, depends on three variables (temperature, volume, fugacity), the pressure and the average number of the particles are  given by
\be
\ba{ll}
P=\kB T \partial_{V}\ln\cZ(T,V,z)\,,&\,
N=z\partial_{z}\ln\cZ(T,V,z)\,,
\ea
\label{Nexpression}
\ee
where $\kB$denotes the Boltzmann constant.  If the system is   finite,  due to the analytic property of the partition function,  the  physical    quantities   which are   given as a fraction between the partition function and its derivatives,  like  (\ref{Nexpression}),  cannot   feature any mathematical singularities.  On the contrary,   infinite systems may do so.  In this way, it seems   that,  \textit{More is the same;  infinitely more is   different}~\cite{Moreisthesame}.

Viewing the Avogadro's number, $\NAv\simeq 6\times 10^{23}$, as an enormous quantity might  well suggest  to take  the infinity limit or the  thermodynamic limit: the  limit of the large volume and the large  number of  particles with the density held fixed~\cite{Huang}.  
Essentially due to the  quantum commutation relation, $[\hat{x},\hat{p}]=i\hbar$, the reduced Planck's constant, $\hbar$,  is  positioned   inside the expression of the  partition function along with  the volume, $V$,  generically through  the combination, $V/\hbar^{3}$, where the power of $\hbar$ corresponds to the dimension of the space.  This implies that  the    large volume limit may  be traded with  the classical limit  $\hbar\rightarrow 0$,  and hence  special care should be taken while considering   the thermodynamic limit, in order to preserve  any quantum nature~\cite{Grossmann95,Kocharovsky2006,KleinertBOOK,GlaumPRA2007}.  Further, since taking the thermodynamic limit and taking the derivatives do not    commute in general,  desirably it is safer to take   the thermodynamic limit only at the end of computation.

Recently two of the authors investigated the isobar of  an ideal Bose gas confined in a  box within the canonical ensemble,  
without assuming the  thermodynamic limit~\cite{7616PRA}.  Numerical computations based on the exact expression of the corresponding canonical partition function revealed that,  if the number of particles is equal to or greater than a certain critical value, which turns out to be $7616$ for the `cubic' box, the isobar zigzags   featuring `$\cS$-shape'  on the $(T,V)$-plane  (\textit{cf.} FIG.~\ref{FIGisobar} in the present paper). The  two turning points on the $\cS$-shaped isobar are naturally  identified as the `{supercooling}' $(T^{\ast},V^{\ast})$ and  the `{superheating}' $(T^{\ast\ast},V^{\ast\ast})$  points.  Between the supercooling and the superheating temperatures, $T^{\ast}<T<T^{\ast\ast}$,  the volume becomes  triple-valued. Since all the physical quantities are functions of the temperature and the volume, every physical quantity itself  is   triple-valued between the two temperatures and  changes discontinuously on isobars as the temperature increases.  In fact, any  temperature derivative  restricted on  isobars diverges  at the points with  the universal  singularity  exponent, $1/2$~\cite{7616relativistic}.  In this way imposing the  `constant pressure constraint,'    a discrete phase transition was for the first time  realized in  a \textit{finite} system,    derived  \textit{ab  initio}    from the corresponding  partition function.

However,    due to the limitation in our computational power (supercomputer, SUN B6048), the  numerical analyses  performed in  Refs.\cite{7616PRA,7616relativistic} were  restricted to the particle numbers not greater than   one million.  In particular, the separation between the supercooling and the superheating temperatures gets wider  as  the number of particles increases within  the range, $7616\leq N\leq 10^{6}$. Hence, it was not  clear, what would happen for much larger number of particles,  or  closer to  the thermodynamic limit.   \newpage

It is the purpose of the present paper,  first    to verify the same feature of the ideal Bose gas within the grand canonical ensemble, both analytically and numerically;  and  second   to   address rigorously  its thermodynamic limiting   behavior.  \\

Basically we set to analyze the following equation~\cite{Kardar} which  shall   be derived  from the grand canonical partition function of the ideal Bose gas:
\be
{\left.\frac{\rd T}{\rd V}\right|_{P,N}=0\,.}
\label{dTdP}
\ee
This condition  is equivalent to the usual definition of the spinodal curve~\cite{Kardar,spinodalnuclear,Sasaki:2007qh,Yukalov},
\be
{\left.\frac{\rd P}{\rd V}\right|_{T,N}=0\,,}
\ee
and  must be met at  the supercooling and the superheating points on isobars.

\section{ANALYSIS}
Essentially due to the  non-relativistic dispersion relation,  $E={\vec{p}{}^{\,2}}/{(2m)}$, where $m$  is  the mass of the particle,   the  grand canonical partition function of the ideal Bose gas  confined in a cubic box  is essentially   a two-variable function depending on  the fugacity, $z$,  and  the combination  of temperature and volume, $TV^{2/3}$.  Specifically  we set,  as for the two  fundamental variables in our analysis,  
\be
\ba{ll}
\varepsilon:= \tfrac{\pi^{2}\hbar^2}{2m\kB}\left(TV^{2/3}\right)^{-1}\,,~~&~~
\sigma:=-\ln z\,.
\ea
\label{fundamental}
\ee
In terms of these, the grand canonical partition function reads
\be
\ln\cZ(\varepsilon,\sigma)=-{\textstyle{\sum}}_{\vn\in\mN^{3}}\,\ln\!\left(1-e^{-\varepsilon\vn^{2}-\sigma}\right)\,.
\label{logZ}
\ee
With the Dirichlet boundary condition which we deliberately  impose, $\vn=(n_1,n_2,n_3)\in\mN^{3}$ is a \textit{positive} integer-valued lattice vector, such that the lowest value of $\vn^{2}$ is the spatial dimension,  $3$, and $\sigma$ is bounded from below
\be
\sigma> -3\varepsilon\,,
\label{sigmabound}
\ee
while  $\varepsilon$ is positive.    Searching  for  spinodal curves near to the thermodynamic limit,  we shall be interested in the small  $\varepsilon$ region.

It is useful to note,  for the computation of various  physical quantities such as  (\ref{Nexpression}),
\be
\ba{ll}
\left.T\partial_{T}\right|_{V,z}=
\tfrac{3}{2}\!\left.V\partial_{V}\right|_{T,z}=-\varepsilon\partial_{\varepsilon}\,,&
\left.z\partial_{z}\right|_{T,V}=-\partial_{\sigma}\,.
\ea
\ee
It follows that the number of particles (\ref{Nexpression}) reads 
\be
N(\varepsilon,\sigma)=-\partial_{\sigma}\ln\cZ(\varepsilon,\sigma)\,,
\label{Nexpression2}
\ee
and the formula of the pressure (\ref{Nexpression}) is equivalent to 
\be
\Tp(\varepsilon,\sigma):=\!\left(\tfrac{2m}{\pi^{2}\hbar^{2}}\right)^{\frac{3}{5}}\!\kB T P^{-\frac{2}{5}}=\left[-\tfrac{2}{3}\varepsilon^{\frac{5}{2}}\partial_{\varepsilon}\ln\cZ(\varepsilon,\sigma)\right]^{-\frac{2}{5}}.
\label{defTp}
\ee
Being a combination of $T$ and $P$,  this  dimensionless quantity, $\Tp$,   can determine  the physical   temperature   on an arbitrarily   given isobar. Similarly we may define  a dimensionless ``volume",
\be
\Vp(\varepsilon,\sigma):=\left(\tfrac{2m}{\pi^{2}\hbar^{2}}P\right)^{\frac{3}{5}}V
=\left[-\tfrac{2}{3}\partial_{\varepsilon}\ln\cZ(\varepsilon,\sigma)\right]^{\frac{3}{5}}\,,
\label{defVp}
\ee
and another dimensionless ``temperature",
\be
\textstyle{\Trho(\varepsilon,\sigma):=\tfrac{2m}{\pi^{2}\hbar^{2}}\kB T\left(\frac{V}{N}\right)^{\frac{2}{3}}
=\left[-\varepsilon^{\frac{3}{2}}\partial_{\sigma}\ln\cZ(\varepsilon,\sigma)\right]^{-\frac{2}{3}}\,.}
\label{defTrho}
\ee
As we   already wrote, $N$, $\Tp$, $\Vp$ and   $\Trho$   are  functions of the two variables, $\varepsilon$, $\sigma$ only. They  satisfy  identities, 
\be
\ba{ll}
\varepsilon\Tp(\varepsilon,\sigma)=\left[\Vp(\varepsilon,\sigma)\right]^{-\frac{2}{3}}\,,~&~
\varepsilon\Trho(\varepsilon,\sigma)=\left[N(\varepsilon,\sigma)\right]^{-\frac{2}{3}}\,.
\ea
\ee

\begin{figure}[h]
\centering\includegraphics[width=81mm]{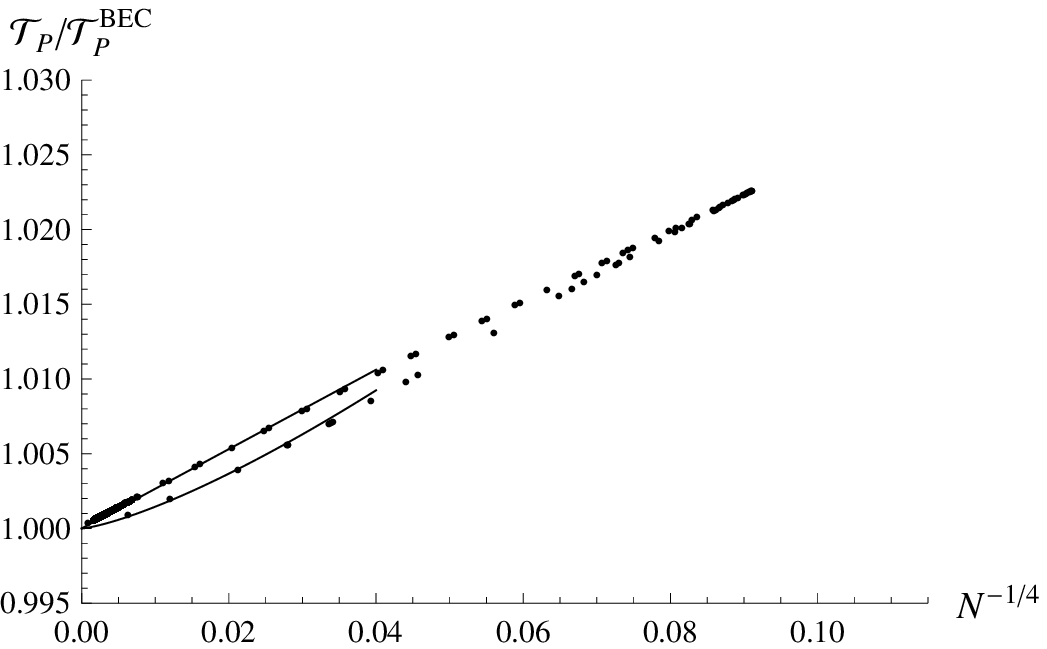}
\caption{The supercooling   and the superheating   spinodal curves  on the $(N^{-1/4},{\Tp/\Tp^{\BEC}})$-plane (lower and upper curves respectively).  
The dotted curves   are from the numerical computations based on the exact formulae (\ref {Nexpression2}), (\ref{defTp}), (\ref{Phispinodal}). The solid lines correspond to our  analytic approximation (\ref{MAINsupercooling}), (\ref{MAINsuperheating}) for large $N$. A pair of spinodal curves start to develop at ${N=N_{\rm{c}}}\simeq 14392.4$ ($N^{-1/4}_{\rm{c}}\simeq 0.0912991$) which is comparable to the critical number  of  the canonical ensemble,  $7616$~\cite{7616PRA}.}
\label{FIGspinodal}
\end{figure}

\begin{figure}[h]
\centering\includegraphics[width=81mm]{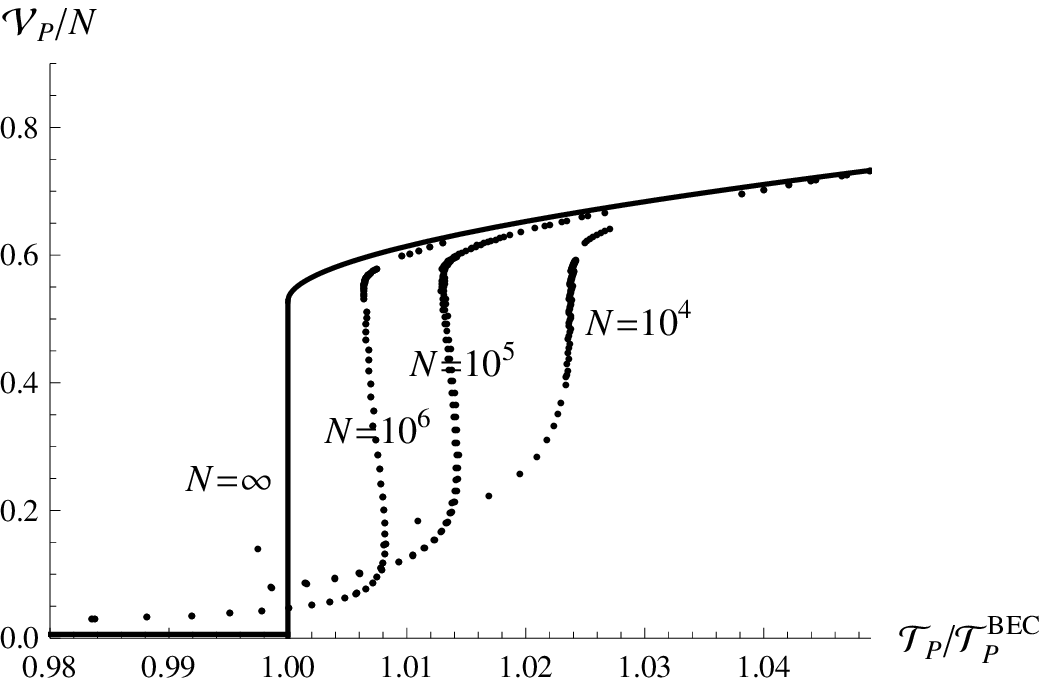}
\caption{Isobar curves on the $({\Tp/\Tp^{\BEC}},{\Vp/N})$-plane. They zigzag featuring `$\cS$-shape'  if $14393\leq N<\infty$. }
\label{FIGisobar}
\end{figure}

Now the spinodal curve (\ref{dTdP})  is positioned on the $(\varepsilon,\sigma)$-plane to satisfy 
${\rd N(\varepsilon,\sigma)=0}$  and ${\rd \Tp(\varepsilon,\sigma)=0}$, such that  the following linear equation must admit   a nontrivial solution,
\be
\left(\ba{c}0\\0\ea\right)=\left(\ba{cc}
\partial_{\varepsilon}\partial_{\sigma}\ln\cZ&\partial_{\sigma}^{2}\ln\cZ\\
\left(\tfrac{5}{2}\varepsilon^{-1}\partial_{\varepsilon}+\partial_{\varepsilon}^{2}\right)\ln\cZ~&\partial_{\varepsilon}\partial_{\sigma}\ln\cZ\ea
\right)\left(\ba{c}\rd \varepsilon\\\rd\sigma\ea\right)\,.
\label{lineareq}
\ee
It follows that the ${2\times 2}$ matrix in (\ref{lineareq}) must be singular,
\be
\Phi:=\det \left(\ba{cc}
\partial_{\varepsilon}\partial_{\sigma}\ln\cZ&\partial_{\sigma}^{2}\ln\cZ\\
\left(\tfrac{5}{2}\varepsilon^{-1}\partial_{\varepsilon}+\partial_{\varepsilon}^{2}\right)\ln\cZ~&\partial_{\varepsilon}\partial_{\sigma}\ln\cZ\ea
\right)\equiv 0\,.
\label{Phispinodal}
\ee
This algebraic equation determines    the spinodal curve  on the $(\varepsilon,\sigma)$-plane.   Further, it is straightforward to show that the determinant  is proportional to   $\left.\frac{\rd \Tp}{\rd \Vp}\right|_{N}$ as
\be
\textstyle{\left.\frac{\rd \ln\Tp}{\rd \ln\Vp}\right|_{N}=\frac{2}{\,3\left(\partial_{\sigma}^{2}\ln\cZ\right)^{2}{\rm Var}(\vn^{2})}\times\Phi\,,}
\ee
where   ${\rm Var}(\vn^{2})$ is our shorthand notation for
\be
\textstyle{{\rm Var}(\vn^{2}):=\frac{\partial_{\varepsilon}^{2}\ln\cZ}{\partial_{\sigma}^{2}\ln\cZ}-\left(\frac{\partial_{\varepsilon}\partial_{\sigma}\ln\cZ}{\partial_{\sigma}^{2}\ln\cZ}\right)^{2}\,,}
\ee
which can be identified  as  the variance of $\vn^{2}$ with respect to the probability  distribution   proportional to  $\sinh^{-2}(\half\varepsilon \vn^{2}+\half\sigma)$~\footnote{In our convention, $\sinh^{-2}(x)=[\sinh(x)]^{-2}$, \textit{etc.}}.  Hence, ${\rm Var}(\vn^{2})$ is positive definite and the vanishing of the determinant  is,  as expected, equivalent to the vanishing of  $\left.\frac{\rd \Tp}{\rd \Vp}\right|_{N}$.  Our main task is to solve  (\ref{Phispinodal}) and  express the solutions in terms of the more physical variables, $N$, $\Tp$,  $\Vp$, $\Trho$   using (\ref{Nexpression2}), (\ref{defTp}), (\ref{defVp}), (\ref{defTrho}).  Our numerical solutions are depicted in FIG.\ref{FIGspinodal} and FIG.\ref{FIGisobar},  along with an analytic approximation which we discuss below.   \\
~\newline
~\newline
\subsection*{Analytic approximation}
Our analytic analysis starts with the following expression for  the  derivatives of the partition function, 
\be
\ba{l}
(\partial_{\varepsilon})^{l}(\partial_{\sigma})^{k-l}\ln\cZ(\varepsilon,\sigma)\\
=\sum_{\vn\in\mN^{3}}\sum_{a=1}^{k}\,(\vn^{2})^{l}\frac{(-1)^{k}C_{k,a}}{\left(e^{\varepsilon \vn^{2}+\sigma}-1\right)^{a}}\\
=\sum_{\vn\in\mN^{3}}\sum_{p=0}^{\infty}\sum_{b=1}^{k+p}\,(\vn^{2})^{l}\left(\frac{\,\sigma^{p}}{p!}\right)\frac{\,(-1)^{k+p}C_{k+p,b}}{\left(e^{\varepsilon \vn^{2}}-1\right)^{b}}\,.
\ea
\label{pcZ}
\ee
Here  $k\geq\max (l,1)$, and $C_{k,a}$ are positive integers that are   determined  by a recurrence relation, 
\be
C_{{k+1},a}=aC_{k,a}+{(a-1)}C_{k,{a-1}}\,,
\label{recr}
\ee
with the initial value, $C_{1,1}=1$. The recurrence relation  comes from the expansion, 
\be
\textstyle{\left(-\frac{\rd~}{\rd x}\right)^{k}}\ln(1-e^{-x})=-\sum_{a=1}^{k}\,\frac{C_{k,a}}{\left(e^{x}-1\right)^{a}}\,.
\label{rrC}
\ee
Taking an  $x$-derivative of the right hand side of the equality in (\ref{rrC}) leads to (\ref{recr}). 
Further, it is  useful to note 
\be
\partial^{2}_{\sigma}\ln\cZ(\varepsilon,\sigma)=\!\sum_{\vn\in\mN^{3}}\!\left[\textstyle{\frac{1}{(\ve\vn^{2}+\sigma)^{2}}}
-\dis{\sum_{k=2}^{\infty}}\textstyle{(\tfrac{1}{4})^{k}\cosh^{-2}\!\left(\frac{\ve\vn^{2}+\sigma}{2^{k}}\right)}
\right].
\label{sigma2Z}
\ee
This expression is  due to an  identity,  
\be
\ba{l}
\sinh^{-2}(x)\\
=(\tfrac{1}{4})^{p}\sinh^{-2}\!\left(\frac{x}{2^{p}}\right)-{\sum_{j=1}^{p}}(\tfrac{1}{4})^{j}\cosh^{-2}\!\left(\frac{x}{2^{j}}\right)\\
=x^{-2}\,-{\sum_{j=1}^{\infty}}(\tfrac{1}{4})^{j}\cosh^{-2}\!\left(\frac{x}{2^{j}}\right)\,,
\ea
\label{idsinh}
\ee
which holds for  an arbitrary positive integer, $p$. Taking $p$ to infinity gives the second  equality in (\ref{idsinh}).\\

In order to compute the sums in (\ref{pcZ}),  we  adopt the following scheme of analytic approximation:
\begin{enumerate}
\item \textit{Introduce a cut-off,} $\Lambda\geq 3$, \textit{for the lattice sum,}
\be
\sum_{\vn\in\mN^{3}}f(\varepsilon\vn^{2})=\sum_{\vn^{2}\leq\Lambda}f(\varepsilon\vn^{2})+\sum_{\vn^{2}>\Lambda}f(\varepsilon\vn^{2})\,.
\label{scheme1}
\ee
\item \textit{Approximate the last term  by an integral,} 
\be
\sum_{\vn\in\mN^{3}}\!f(\varepsilon\vn^{2})\simeq\!\sum_{\vn^{2}\leq\Lambda}\!f(\varepsilon\vn^{2})+\!\int_{\varepsilon\Lambda}^{\infty}\!\rd x\,
(\textstyle{\frac{\pi}{4}\varepsilon^{-\frac{3}{2}}x^{\frac{1}{2}}-\frac{\,3\pi}{8}\varepsilon^{-1}})f(x).
\label{schemeMain}
\ee
\item   \textit{Put}  $\sigma=-\varepsilon\mu$ \textit{with  a new variable}, $\mu$. \textit{From} (\ref{sigmabound}), $\mu<3$.  
\item \textit{Keep only the dominant singular terms in the power series  expansion of}   (\ref{schemeMain}) \textit{in} $\varepsilon$, \textit{which  are manifestly  cut-off independent.  Allow ${{\mu}}$ to be  expandable in $\ve$ with an arbitrary leading power.}
\end{enumerate}
The approximation (\ref{schemeMain}) can be traced back to an identity, 
\be
\ba{ll}
\sum_{\vn\in\mN^{3}}f(\varepsilon\vn^{2})=&\tfrac{1}{8}\!\left[\sum_{\vn\in\mZ^{3}}f(\varepsilon\vn^{2})\right]-
\tfrac{3}{8}\!\left[\sum_{\vn\in\mZ^{2}}f(\varepsilon\vn^{2})\right]\\
{}&+\tfrac{3}{8}\!\left[\sum_{n\in\mZ}f(\varepsilon n^{2})\right]-\tfrac{1}{8}f(0)\,,
\ea
\label{lsum}
\ee
where  the first two sums on the right hand side of the equality can  be approximated  by integrals in  three or  two dimensional spherical coordinates,  and  the remaining part may be neglected  for small $\ve$~(see \cite{Grossmann95} and references therein).\\

With the constants, 
\be
\ba{ll}
{a_{s}:={\int_{0}^{\infty}\rd x}
{\frac{x^{s}}{e^{x}-1}}=\Gamma(s+1)\zeta(s+1)\,,}\\
\textstyle{b:={\int_{0}^{\infty}\rd x\,}\sqrt{x}\cosh^{-2}(x)\,,}
\ea
\ee
and   the estimations~\cite{Grossmann95},
\be
\textstyle{\int_{\ve \Lambda}^{\infty}\rd x\,\frac{1}{e^{x}-1}\simeq\int_{\ve \Lambda}^{\infty}\rd x\,\frac{xe^{x}}{(e^{x}-1)^{2}}
\simeq -\ln\ve\,,}
\ee
our scheme enables us to compute 
\be
\ba{l}
\partial_{\varepsilon}\ln\cZ\simeq-\langle\frac{\,3\ve^{-1}}{3-\mu}\rangle_{-2}-\tfrac{\pi}{4}a_{\frac{3}{2}}\varepsilon^{-\frac{5}{2}}+\tfrac{3\pi}{8}a_{1}\varepsilon^{-2}\,,\\
\partial_{\sigma}\ln\cZ\simeq-\langle\frac{\,\ve^{-1}}{3-\mu}\rangle_{-\frac{3}{2}}
-\tfrac{\pi}{4}a_{\frac{1}{2}}\varepsilon^{-\frac{3}{2}}-\tfrac{3\pi}{8}\ve^{-1}\ln\ve\,,\\
\partial_{\varepsilon}^{2}\ln\cZ\simeq
\langle\frac{9\ve^{-2}}{(3-\mu)^{2}}\rangle_{{-3}}+
\tfrac{5\pi}{8}a_{\frac{3}{2}}\varepsilon^{-\frac{7}{2}}-\tfrac{3\pi}{4}a_{1}\varepsilon^{-3}\,,\\
\partial_{\varepsilon}\partial_{\sigma}\ln\cZ\simeq
\langle\frac{3\ve^{-2}}{(3-\mu)^{2}}\rangle_{{-\frac{5}{2}}}+\tfrac{3\pi}{8}a_{\frac{1}{2}}\varepsilon^{-\frac{5}{2}}+\tfrac{3\pi}{8}\varepsilon^{-2}{\ln\ve}\,,\\
\partial_{\sigma}^{2}\ln\cZ\simeq\frac{\ve^{-2}}{(3-\mu)^{2}}+\left[\sum_{\vn^{2}>3}\frac{1}{(\vn^{2}-\mu)^{2}}\right]\!\ve^{-2}-\frac{\,(2+\sqrt{2})\pi}{8}b\,\ve^{-\frac{3}{2}}\,,
\ea
\label{pcZ2}
\ee
where  $\langle \,g(\ve)\,\rangle_{-n}\!$ denotes a  part of the series expansion of $g(\ve)$ in $\ve$ which is at least $(-n)$-th order singular, for example,
\be
\ba{ll}
\langle \ve^{-\frac{3}{2}}+\ve^{-1}+1+\ve\rangle_{-\frac{3}{2}}=\ve^{-\frac{3}{2}}\,,&{}\\
\langle \ve^{-\frac{3}{2}}+\ve^{-1}+1+\ve\rangle_{-1}=\ve^{-\frac{3}{2}}+\ve^{-1}\,,&\mbox{~etc.}
\ea
\ee
Especially for $\partial_{\sigma}^{2}\ln\cZ$, it is important  to note that the sum, $\sum_{\vn^{2}>3}(\vn^{2}-\mu)^{-2}$,    converges,  since   
\be
\ba{lll}
\sum_{\vn^{2}>\Lambda}\frac{1}{(\vn^{2}-\mu)^{2}}&\leq&\sum_{\vn^{2}>\Lambda}\frac{1}{(\vn^{2}-|\mu|)^{2}}\\
{}&\simeq&\int_{\Lambda}^{\infty}\!\rd x\,
\frac{\,\frac{\pi}{4}x^{\frac{1}{2}}-\frac{\,3\pi}{8}}{(x-|\mu|)^{2}}\\
{}&=&\textstyle{
\tfrac{\pi}{8}\left[\frac{\,2\sqrt{\Lambda}-3\,}{\Lambda-|\mu|}+
\frac{1}{\sqrt{|\mu|}}\ln\left(\frac{\,\sqrt{\Lambda}+\sqrt{|\mu|}\,}{
\sqrt{\Lambda}-\sqrt{|\mu|}}\right)\right]}\,.
\ea
\label{sinh2}
\ee
The numerical values of the constants are
\be
\ba{ll}
a_{\frac{1}{2}}=\tfrac{\sqrt{\pi}}{2}\zeta(\tfrac{3}{2})\simeq 2.31516\,,~~&~a_{1}=\tfrac{\,\pi^{2}}{6}\simeq 1.64493\,,\\
a_{\frac{3}{2}}=\tfrac{\,3\sqrt{\pi}}{4}\zeta(\tfrac{5}{2})\simeq  1.78329\,,~~&~b\simeq 0.758128\,.
\ea
\ee
Having the expressions (\ref{pcZ2}),    we now proceed to solve the spinodal curve condition (\ref{Phispinodal}).  Since the indices, $n$ of the symbol, $\langle~\cdot~\rangle_{-n}$ appearing in (\ref{pcZ2}) are various,  letting  the leading singular term of  $\frac{\ve^{-1}}{3-\mu}$ be order of $\ve^{-h}$, we need to  separately  consider   the following nine possible  cases: 
\[
\ba{l}
{h<1\,,~~~h=1\,,~~~ 1< h<\frac{5}{4}\,,~~~ h=\frac{5}{4}\,,~~~\frac{5}{4}< h<\frac{3}{2}}\,,\\
{h=\frac{3}{2}\,,~~~ \frac{3}{2}<h<2\,,~~~ h=2\,,~~~2<h}\,.
\ea
\]
Keeping only the two dominant terms in (\ref{pcZ2}) for each case,  it is straightforward to check  that only  the two cases, ${h=1}$ and ${h=2}$, admit solutions, and hence there are  two spinodal curves as follows.\\

{${\mathbf{\bullet}}$\,} {On the $(\ve,\mu)$-plane.}
\begin{list}{$\ast$}{}
\item constant $\mu\simeq\mu^{\ast}$ line with  ${h=1}$,  satisfying 
\be
\textstyle{\sum_{\vn\in\mN^{3}}\frac{1}{(\vn^{2}-\mu^{\ast})^{2}}=\tfrac{9}{8}\left[\zeta(\tfrac{3}{2})\right]^{2}\,.}
\ee
Numerically we get 
\be
\mu^{\ast}\simeq 2.61873\,.
\label{mustar}
\ee
\item$\!\!\ast$  Linear line with  ${h=2}$,
\be
\mu\simeq\mu^{\ast\ast}(\ve)= 3-\tfrac{240}{\pi^{3}}\,\ve\,.
\label{superheatingvemu}
\ee
\end{list}

{${\mathbf{\bullet}}$\,} {In terms of the physical variables,  $N$, $\Tp$, $\Vp$, $\Trho$}. 
\begin{list}{$\ast$}{}
\item {\it{Supercooling spinodal curve}}, for ${h=1}$,
\be
\ba{l}
\Tp^{\ast}/{\Tp^{\BEC}}\simeq 1+\textstyle{{\frac{\,\pi^{3}}{60}\left[\left(\Tp^{\BEC}\right)^{5}/\Trho^{\BEC}\right]^{\frac{1}{2}}}}\,N^{-\frac{1}{3}}\,,\\
\Vp^{\ast}\simeq \textstyle{\left({\Trho^{\BEC}}/{\Tp^{\BEC}}\right)^{\frac{3}{2}}\left(N+\tfrac{\pi}{4}\Trho^{\BEC}\,N^{\frac{2}{3}}\ln N\right)}\,,\\
\Trho^{\ast}/{\Trho^{\BEC}}\simeq 1+\tfrac{\pi}{6}\Trho^{\BEC}\,N^{-\frac{1}{3}}\ln N\,.
\ea
\label{MAINsupercooling}
\ee
\item$\!\!\ast$ {\it{Superheating spinodal curve}},  for ${h=2}$,
\be
\ba{l}
\Tp^{\ast\ast}/{\Tp^{\BEC}}\simeq 1+\tfrac{1}{150}\left(\tfrac{\,\pi^{15}}{15}\right)^{\frac{1}{4}}
\left(\Tp^{\BEC}\right)^{\frac{5}{2}}\,N^{-\frac{1}{4}}\,,\\
\Vp^{\ast\ast}\simeq  8\left(\tfrac{15}{\,\pi^{3}}\right)^{\frac{3}{4}}
\left(\Tp^{\BEC}\right)^{-\frac{3}{2}}\,N^{\frac{3}{4}}\,,\\
\Trho^{\ast\ast}\simeq 4\left(\tfrac{15}{\,\pi^{3}}\right)^{\frac{1}{2}}\,N^{-\frac{1}{6}}\,.
\ea
\label{MAINsuperheating}
\ee
\end{list}
In the above, $\Tp^{\BEC}$ and $\Trho^{\BEC}$ denote two  constants, 
\be
\ba{l}
\textstyle{
\Tp^{\BEC}=\left(\tfrac{64}{\,\pi^{3}}\right)^{\frac{1}{5}}\left[\zeta(\frac{5}{2})\right]^{-\frac{2}{5}}}\simeq 1.02781\,,\\
\textstyle{
\Trho^{\BEC}=\tfrac{4}{\pi}\left[\zeta(\frac{3}{2})\right]^{-\frac{2}{3}}\simeq 0.671253\,,}
\ea
\label{BECtemp}
\ee
which correspond   to  the  well-known Bose-Einstein condensation temperatures for the variables,  $\Tp$ (\ref{defTp}) and $\Trho$ (\ref{defTrho}), the definitions of which we recall  here,
\be
\ba{ll}
\Tp:=\left(\tfrac{2m}{\pi^{2}\hbar^{2}}\right)^{\frac{3}{5}}\!\kB T P^{-\frac{2}{5}}\,,~~&~~\Trho:=\tfrac{2m}{\pi^{2}\hbar^{2}}\kB T\left(\frac{V}{N}\right)^{\frac{2}{3}}\,,\\
\multicolumn{2}{c}{\Vp:=\left(\tfrac{2m}{\pi^{2}\hbar^{2}}P\right)^{\frac{3}{5}}V=
N\left(\frac{\Trho}{\Tp}\right)^{\frac{3}{2}}\,.}
\ea
\ee

\section{DISCUSSION} 
As computable from our analytic expressions,  (\ref{MAINsupercooling}) and (\ref{MAINsuperheating}),  the separation between the supercooling and the superheating temperatures becomes maximal,  if the number of particle is equal to 
\be
N_{\mathrm{{\scriptscriptstyle{MAX}}}}=\textstyle{\frac{5^{15}}{(27\pi)^{3}}\left[\zeta(\frac{3}{2})\right]^{4}\simeq 2.32890\times 10^{6}}\,.
\ee
This also  agrees with the numerical result in FIG.~\ref{FIGspinodal}, as $(N_{\mathrm{{\scriptscriptstyle{MAX}}}})^{-1/4}\simeq 0.0255984$.   When  the number of particles exceeds   this critical value,  the two temperatures,  $\Tp^{\ast}$ and $\Tp^{\ast\ast}$, -- satisfying  $\Tp^{\BEC}<\Tp^{\ast}<\Tp^{\ast\ast}$ --  gets closer, and eventually converges to the   BEC temperature, $\Tp^{\BEC}$ (\ref{BECtemp}),  in the thermodynamic limit. That is to say,   $N_{\mathrm{{\scriptscriptstyle{MAX}}}}$ is the critical number for the thermodynamic limit to work.\\
\indent The   ratio of the two volumes,  
\be
\Vp^{\ast}/\Vp^{\ast\ast}\simeq \left(\tfrac{\pi}{15}\right)^{\frac{3}{4}}\textstyle{\left[\zeta(\frac{3}{2})\right]^{-1}}N^{\frac{1}{4}}\simeq 0.118511\times N^{\frac{1}{4}}\,,
\label{VolumeExp}
\ee
enables us to estimate the discrete volume expansion rate at the liquid-gas type phase transition.   For the Avogadro's number, $\NAv\simeq 6.02214\times 10^{23}$, the volume expansion rate (\ref{VolumeExp}) gives      $\Vp^{\ast}/\Vp^{\ast\ast}\simeq 104399$.  Thus,  the ideal Bose gas made up of the   Avogadro's number of particles   expands its volume  discretely about $10^{5}$ times during  the phase transition.  This is a genuine \textit{finite effect} of the Avogadro's number, which cannot be seen directly  in the thermodynamic limit where $\Vp^{\ast}/\Vp^{\ast\ast}\rightarrow\infty$. \\
\indent Our numerical computations based on the exact formulae   quantitatively agree  with  the  canonical ensemble  results~\cite{7616PRA}  for ${N=10^{5}}$ and ${N=10^{6}}$ within $0.1\%$ error, though the minimum (natural) numbers required for the emergence of  the spinodal curves are different, $14393$ \textit{vs.} $7616$.
\vfill
\begin{table}[H]
\begin{center}
\begin{tabular}{c|c|c}
$(\Tp^{\ast},\Tp^{\ast\ast})$&~~Grand canonical~~&~~Canonical~~\\
\hline
$N=10^{5}$&~$(1.041,1.043)$~&~$(1.0410, 1.0424)$~\\
$N=10^{6}$&~$(1.0348, 1.0364)$~&~$(1.034, 1.036)$
\end{tabular}
\caption{Quantitative agreement between the canonical and the grand canonical  results, within $0.1\%$ error.}
\label{tableCOMPARE}
\end{center}
\end{table}
\vfill
In this work, we have focused on the Dirichlet   boundary condition. Alternatively imposing  periodic or Neumann boundary condition brings out a volume independent ground state energy which, as shown in \cite{7616PRA},  causes a thermodynamic instability at low temperature near absolute zero (see also \cite{differentBC}).  This further implies that,   under the  alternative  boundary conditions,  periodic or Neumann,  the isobar on the $(\Tp,\Vp/N)$ plane is of `$\cC$-shape', rather than of the zigzagging   `$\cS$-shape' as in FIG.\ref{FIGisobar}: Namely  there is a nontrivial   lower bound  in  $\Tp$ of the isobar,  above which the volume is always  double-valued. In the thermodynamic limit,  the lower bound  converges to $\Tp^{\BEC}$,  and the isobar eventually becomes  independent of the boundary conditions, identical to the case of  ${N=\infty}$ in FIG.\ref{FIGisobar}, except for  ${\Vp/N=0}$. When  ${\Vp/N=0}$, under  the periodic or Neumann boundary condition,  $\Tp$ may assume any value which is  greater than or equal to $\Tp^{\BEC}$ (as anticipated in  Fig.\,12.8 of \cite{Huang}),  while under   the Dirichlet   boundary condition, it is quite the opposite,  $0\leq\Tp\leq\Tp^{\BEC}$,  as depicted in  FIG.\ref{FIGisobar}. \\

\indent In conclusion,    we have shown, both numerically and analytically,   that the isobar of the ideal Bose gas  zigzags on the temperature-volume plane,  qualitatively  featuring   the liquid-gas transition,    if $N\geq 14393$. This is   an emergent  phenomenon of the finitely many  bosonic identical particles.    We have  derived  the precise formulae for the  the two turning points: supercooling (\ref{MAINsupercooling}) and superheating (\ref{MAINsuperheating}). Our formulae  reveal  an $N^{-{1/3}}$ or $N^{-{1/4}}$ power  correction to the BEC temperature and enable us to estimate the volume expansion rate, (\ref{VolumeExp}).     \\

\begin{center}
\textbf{Acknowledgement} 
\end{center}
\indent We  thank   Konstantin Glaum, Petr Jizba,   Hagen Kleinert  and Hyun-Woo Lee for helpful  comments.  
SWK  was supported by a grant-in-aid from the Japanese Ministry of Education, Culture, Sports, Science and Technology (No. 20105002).
JHP was supported by the National Research Foundation of Korea(NRF) grant funded by the Korea government(MEST) with the Grant No. 2005-0049409 (CQUeST) and No. 2010-0002980.
\vfill

\end{document}